\documentclass{svjour3}
\smartqed  
\usepackage{graphicx,color}
\usepackage{fix-cm}

\begin{document}


\title{An interleaved sampling scheme for the characterization of single qubit dynamics}

\author{Jason F. Ralph \and Joshua Combes \and Howard M. Wiseman}
\institute{J.F.Ralph \at Department of Electrical Engineering and Electronics, The University of Liverpool, Brownlow Hill, 
Liverpool, L69 3GJ, United Kingdom. \and J.Combes \at Centre for Quantum Dynamics, Griffith University, Brisbane, Queensland 4111, Australia.
\and H.M.Wiseman \at  Centre for Quantum Dynamics, Griffith University, Brisbane, Queensland 4111, Australia.}

\date{\today}
\newcommand{\red}{\color{red}}
\newcommand{\beq}{\begin{equation}}
\newcommand{\eeq}{\end{equation}}
\newcommand{\beqa}{\begin{eqnarray}}
\newcommand{\eeqan}{\end{eqnarray*}}
\newcommand{\beqan}{\begin{eqnarray*}}
\newcommand{\eeqa}{\end{eqnarray}}
\newcommand{\bra}[1]{\langle{#1}|}
\newcommand{\Bra}[1]{\left\langle{#1}\right|}
\newcommand{\ket}[1]{|{#1}\rangle}
\newcommand{\Ket}[1]{\left|{#1}\right\rangle}
\newcommand{\ip}[1]{\langle{#1}\rangle}
\newcommand{\Ip}[1]{\left\langle{#1}\right\rangle}
\newcommand{\non}{\nonumber}
\newcommand{\id}{\openone}
\newcommand{\eq}[1]{Eq.~(#1)}
\newcommand{\eqr}[1]{Eq.~(\ref{#1})}
\newcommand{\eqrs}[1]{Eqs.~(#1)}
\newtheorem{theo}{Theorem}

\newcommand{\bqa}{\begin{eqnarray}}
\newcommand{\eqa}{\end{eqnarray}}
\newcommand{\nn}{\nonumber}
\newcommand{\nl}[1]{\nn \\ && {#1}\,}
\newcommand{\erf}[1]{Eq.~(\ref{#1})}
\newcommand{\erfs}[2]{Eqs.~(\ref{#1})--(\ref{#2})}
\newcommand{\dg}{^\dagger}
\newcommand{\rt}[1]{\sqrt{#1}\,}
\newcommand{\smallfrac}[2]{\mbox{$\frac{#1}{#2}$}}
\newcommand{\half}{\smallfrac{1}{2}}
\newcommand{\ito}{It\^o }
\newcommand{\str}{Stratonovich }
\newcommand{\du}{\partial}
\newcommand{\dbd}[1]{{\du}/{\du {#1}}}
\newcommand{\sq}[1]{\left[ {#1} \right]}
\newcommand{\cu}[1]{\left\{ {#1} \right\}}
\newcommand{\ro}[1]{\left( {#1} \right)}
\newcommand{\an}[1]{\left\langle{#1}\right\rangle}
\newcommand{\st}[1]{\left|{#1}\right|}
\newcommand{\s}[1]{\hat \sigma_{#1}}
\newcommand{\artanh}{{\rm artanh}}

\maketitle

\begin{abstract} 
In this paper, we demonstrate that interleaved sampling techniques can be used to characterize the Hamiltonian of a qubit and its environmental decoherence rate. The technique offers a significant advantage in terms of the number of measurements that are required to characterize a qubit. When compared to the standard Nyquist-Shannon sampling rate, the saving in the total measurement time for the interleaved method is approximately proportional to the ratio of the sample rates. 
\end{abstract}




\section{\label{sec:sec1}INTRODUCTION}

\noindent
The qubit is the principal building block for most quantum information processing schemes. The ability to manipulate the quantum state of one or more qubits to perform useful quantum computing operations is reliant on a number of factors. In this paper we consider two: the estimation of the qubit Hamiltonian and the characterization of environmental decoherence via a sequence of projective measurements. The use of projective measurements to characterize a qubit is not new, it has been demonstrated by a number of groups for single qubits and two qubits (e.g.~\cite{Nakamura99,Pashkin03a,Duty04}) and methods for interpreting the measurements to reconstruct the Hamiltonian and the environmental decoherence rate have been considered (e.g.~\cite{Schirmer04,Cole05,Cole06}). 

In this paper, we propose a characterization method based on classical sampling methods which could be a more efficient way to determine the dynamical parameters of an experimental qubit. Specifically, we demonstrate that, when the decoherence in the qubit is low enough for the frequency components of the coherent oscillations to be confined to a relatively small band of frequency space, signal reconstruction techniques based on two sets of interleaved measurements provide robust estimates of the relevant parameters with significantly fewer measurements. The reduction in the number of measurements is very important in this case because the use of projective measurements means that the system is reinitialized in an eigenstate of the measured quantity after each measurement the total time needed to take the measurements scales as the square of the number of measurements, ignoring the time required for read-out and re-initialisation of the qubit. 

We begin by introducing the interleaved sampling (or signal reconstruction) technique in section \ref{sec:sec2}. Then in section \ref{sec:sec3} we show how it can be applied to a signal that is generated by a series of projective measurements. Also we describe the qubit model that we use to illustrate the benefits of interleaved reconstruction. Section \ref{sec:sec4} presents examples of the reconstruction method and the demonstrates the superior accuracy of the interleaved sampling method for estimating the decoherence rate of the qubit. In section \ref{sec:sec5} we conclude. 

\section{\label{sec:sec2}INTERLEAVED SAMPLING}

\noindent
The standard approach to the reconstruction of classical signals is to use the Nyquist-Shannon sampling theorem to determine the minimum sampling rate. If a signal, $x(t)$, has no frequency components above a finite cut-off  frequency ($B$), then the signal can be reconstructed exactly by sampling at a rate of $f_S= 2B$ ~\cite{Jerri77} and then interpolating the signal between the sample points using normalized sinc functions~\cite{Jerri77}. For a series of samples $x[n]$ (where $n$ is an integer, and $x[n]$ corresponds to a measurement at time $t_n = n\Delta t = n/2B$), the interpolated function is given by
\begin{equation}\label{sample1}
x(t) = \sum_{n=-\infty}^{\infty}x[n]{\rm sinc}\left(\frac{t- t_n}{\Delta t} \right).
\end{equation}
The reconstructed signal is an exact reconstruction of $x(t)$ provided that $x(t)$ is limited to the band $[0,B]$. In the frequency domain the sampling process causes the reconstructed signal to be copied into frequency bands at integer multiples of $f_{S}$; because the signal is band-limited the original spectrum does not overlap with these copies.

In practice, real signals are not limited to a finite band of frequencies. The frequency transform of a causal signal can be zero at discrete values but not over any finite band~\cite{Lathi98}. Thus the signal will not be completely confined to the required frequency band. When the signal is not band-limited, significant portions of the spectrum outside the relevant frequency band will be copied into the band of interest during the sampling process. This is referred to as aliasing. However, as long as the signal components outside the band are negligible, the distortion introduced by aliasing will be small.

The minimum sampling rate of $f_S = 2B$ and the reconstruction formula, \erf{sample1}, are applicable to `baseband' signals (i.e. ones that contain low frequency components near DC and up to some maximum frequency $B$). For signals that do not contain significant components at low frequencies, it is possible to reduce the sampling rate by generating two lower-rate measurement series and then use a more complex reconstruction formula, due to Kohlenberg~\cite{Kohlenberg}, which interleaves the two series of measurements. If the signal lies between a lower frequency $f_L$ and an upper frequency such that $f_U = f_L+B$, where the signal bandwidth is $B$, the two sample rates for the interleaved sampling need only be $f_S = B$. That is, the total sample rate is $2B$ rather than the $f_S = 2(f_L+B)$ that would be required if the standard interpolation/reconstruction formula were to be used. If $f_L \gg B$, the saving in measurements is considerable.

The two sampling processes are taken at intervals $\Delta t = 1/B$ and separated by a time $k$. The separation between samples, $k$, can take any value, $0 < k < \Delta t$, as long as the interpolation function is well-behaved. The generalized interpolation formula is given by~\cite{Vaughan91},
\begin{equation}
x(t) = \sum_{n=-\infty}^{\infty}
\left[\begin{array}{l}
x_0[n] S(t-t_n) +x_k[n]S(-t+t_{n,k})
\end{array}\right]\label{sample2}
\end{equation}
where the two series $x_0[n]$ and $x_k[n]$ are measurements at $t_n=n\Delta t$ and $t_{n,k}=(n\Delta t+k)$ respectively, and the interpolation function $S(t)$ is
\begin{equation}
S(t) = S_0(t)+S_1(t)
\end{equation}
where
\begin{eqnarray}
S_0(t) &=& \frac{\cos[2\pi(rB-f_L)t-r\pi B k]-\cos[2\pi f_L t-r\pi B k]}
{2\pi B t \sin(r\pi B k)},\\
\nonumber S_1(t) &=& \frac{
\cos[2\pi(f_L+B)t-(r+1)\pi B k] -\cos[2\pi (rB-f_L) t-(r+1)\pi B k] }
{2\pi B t \sin[(r+1)\pi B k ]},\\
\end{eqnarray}
where $r$ is {the smallest integer larger than $\frac{2f_L}{B}$. The interpolation function has the values: $S(0) = 1$ and $S(n\Delta t) = S(n\Delta t+k) = 0$.   For the special case where the lower frequency $f_L$ is zero and $k = \Delta t/2 = 1/2B$, the interpolation formula reduces to (\ref{sample1}).

We now consider how this interleaved sampling method could be used to characterize the Hamiltonian and decoherence properties of a qubit. 

\section{\label{sec:sec3}QUBIT MODEL AND MEASUREMENT}

\noindent
There are several key differences between the reconstruction of a classical signal and the reconstruction of quantum coherent oscillations. The primary difference is the unavoidable backaction in quantum measurements. Quantum systems cannot be continuously measured without disturbing the dynamics one is trying to probe. Continuous measurements are possible, however continuous quantum parameter and state estimation can be computationally expensive, see Refs.~\cite{Ralph09} and \cite{Gambetta01}. It is for this reason that most treatments are restricted to periods of evolution punctuated by projective measurements of an observable $\sigma_{z}$ for example. Such measurements only ever provide a $+1$ or $-1$ measurement result, with the probability of the result $\pm1$ determined by the quantum state. The results $\pm 1$ correspond to the $\ket{+}=(1,0)^T$ and $\ket{-}=(0,1)^T$ eigenstate of the measured quantity. The sample values that we need to reconstruct the signal are found by taking a series of measurements -- which is termed the record (one realization of the random process) -- and averaging the record over multiple realizations. There are $N$ measurements at each sample point in the record, giving a measurement error (projection noise) which scales as $1/\sqrt{N}$. In the limit $N\rightarrow \infty$, the projection noise vanishes. After each measurement the quantum system has been projected into one or other of the eigenstates -- effectively reinitializing the system -- and the evolution/oscillations must begin again. This means that the total time taken for a series of $M$ sample points is at least
\begin{equation}
T_{\rm min} = N (\Delta t+2\Delta t \ldots + M\Delta t) 
=\smallfrac{1}{2}N M (M+1)\Delta t
\end{equation}
so the total time taken to generate $M$ sample points scales approximately as the square of $M$, and any reduction in the number of points has a large effect on the minimum time taken to characterize the qubit. However, the time interval $\Delta t$ is longer for the interleaved sampling method because the interleaved sample rate is lower than the Nyquist-Shannon sampling rate, so the saving in the total measurement time is approximately equal to the ratio of the sample frequencies. It is worth noting that there is also a saving in the time required to generate the measurement record by not re-preparing the initial ($+1$) eigenstate after each measurement if the $-1$ result is obtained. This can be achieved in software for both the sinc and interleaved reconstruction methods.

The state of the qubit is represented by the density matrix $\rho$ and its evolution can be found by solving the master equation. For a Hermitian Lindblad operator $\hat{c} = \sqrt{\kappa}\hat{y} = \hat{c}^{\dagger}$, where $\kappa$ is the strength of the environmental interaction, this is given by~\cite{WisMil10}  
\begin{equation}\label{me}
\dot{\rho}
=-i\Big[\hat{H},\rho\Big]-\frac{\kappa}{2}\left[\hat{y}\left[\hat{y},\rho\right]  \right].
\end{equation}
The typical decoherence timescale is $\tau = 1/\kappa$. This equation assumes that the environmental noise is uncorrelated (Markovian) and that the coupling to the environment is relatively weak.

We can rewrite the Master equation in the Bloch vector representation~\cite{WisMil10}, with
\begin{equation}
r_i = {\rm Tr}[\sigma_i \rho] \hspace{5mm} i \in  \{x,y,z\} 
\end{equation}
where $\sigma_i$ is a Pauli matix and the density operator is
\begin{equation}
\rho = \smallfrac{1}{2}{(I+r_x\sigma_x+r_y\sigma_y+r_z\sigma_z)}. 
\end{equation}
With a Hamiltonian $H=\omega\sigma_x/2$ (where $\omega = 2\pi f$ and $f$ is the characteristic qubit oscillation frequency) and $\hat{y} = \sigma_z$, equation (\ref{me}) provides three coupled equations
\begin{eqnarray}\label{me4}
\dot{r}_x &=&  - 2\kappa r_x , \nonumber \\
\dot{r}_y &=&  -\omega r_z - 2\kappa r_y , \\
\dot{r}_z&=&  \omega r_y  . \nonumber
\end{eqnarray}
which can be solved analytically. For an initial condition $r_{z}=1$ and $r_x = 0$,
\begin{eqnarray}
r_x(t) &=& 0 \label{me1}\\
r_y(t) &=&  -\frac{\omega}{ \mu} e^{-2 \kappa t} \sin (\mu t) \label{me2} \\
r_z(t) &=& e^{-2 t \kappa }  \left[
\cos(t \mu)
+{2 \kappa  } \;{\sin(t
\mu)}/ \mu\right] \label{me3}
\end{eqnarray} 
where $\mu = \sqrt{\omega ^2-4 \kappa ^2}$. Restricting ourselves to a measurement of the $z$ component of spin, for simplicity, the probability of getting a $+1$ measurement result at a time $t$ is simply $P_+(t) = (r_z+1)/2$ and $P_-(t) = 1-P_+(t)$ for the $-1$ result. We can use the analytic expressions to find the probabilities for the result of each measurement as a function of time and generate a series of $N$ measurements to simulate a set of experiments. By generating a set of $N$ results at each of the sample points, we can average the results to find an estimate the signal level/oscillation at the appropriate times and then use these values in the interpolation/reconstruction formula (\ref{sample2}) as if it were a noisy classical signal.

The Hamiltonian and measurement were selected as an example because they represent a common solid state device: a charge qubit quantum dot with a tunneling interaction and a charge measurement~\cite{Nakamura99,Pashkin03a,Duty04,WisMil10}.

\section{\label{sec:sec4}RESULTS}

\noindent Figure~\ref{fig:Fig1} shows an example of the reconstruction of coherent oscillations, that is \erf{me3}, using both the sinc [see Fig.~\ref{fig:Fig1} (a)] and interleaving methods [see Fig.~\ref{fig:Fig1} (b)]. The value of $\kappa$ selected is at the upper end of the range for which the technique can be used (in practice $\kappa \le 0.3$). The lower the value of $\kappa$, the narrower the spectral profile and the better the approximation that the signal is confined to the selected frequency band. Ideally, the signal outside the frequency band should be as low as possible, but there is a trade-off between widening the band and reducing the number of measurements used in the interleaving. If the decoherence is sufficiently weak for the frequency components outside the band $[f_L,f_U]$ to be small, then the interleaving technique is remarkably robust. From figure~\ref{fig:Fig1} it is apparent that the sinc reconstruction is accurate and appears to be robust to projection noise.  This must be evaluated with regard to our aims of time-efficiently characterizing the fundamental frequency and the decoherence parameter. With this in mind consider the following: the sinc method is using 4200 samples compared to 1800 samples for the interleaving method. In this example, the number of samples for the interleaved method is approximately $0.43$ times the number of samples required by the standard sinc method.

\begin{figure}
	\includegraphics[height=0.55\hsize]{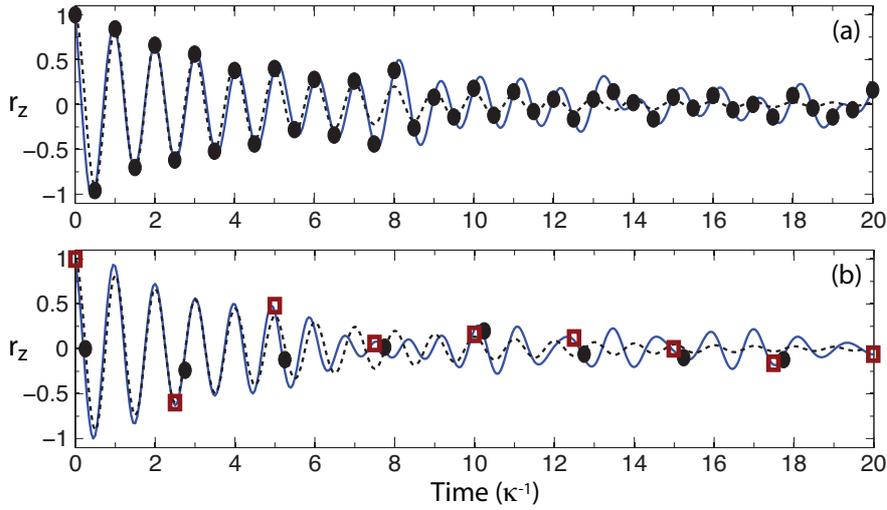}
	\caption{\label{fig:Fig1} (Color online) Signal reconstruction with $\kappa =0.1$, $f=1$, $f_{L}=0.8$, $B=0.4003$. The dashed lines are a plot of \erf{me3}. (a) the solid blue line is the sinc reconstruction, i.e. \erf{sample1} of coherent oscillations from the noisy signal from $N=100$ projective measurements per point ($M=42$, two samples not shown). The solid circles represent the averaged record, which is plotted at times $1/f_{S}$  (b) the squares and circles are the two interleaved (averaged) records with $N=100$ ($M=18$, one sample not shown). The solid blue line is the interleaved reconstruction, ie. \erf{sample2}.}
\end{figure}
Figure~2 shows the Fourier amplitude of the frequency components in and around a frequency band $f_L = 0.8$ to $f_U = f_L+B = 1.2$ for a central frequency of $f= 1$, $\kappa = 0.02$ ($\tau \approx 50$), $N=100$, and $M=160$ (over $200$ oscillation cycles). The number of measurements required for the interleaved method is three times smaller than that required from Nyquist-Shannon sampling, and the minimum time required to take the measurements is reduced by a similar factor ($\times 3.03$). The central peak in the frequency band is clearly visible and the figure also shows a Lorentzian resonance curve that has been fitted to the data in the frequency band. The main frequency components are present in the band $[f_L, f_U]$. Figure 2 also depicts the (noisy) tails of the distribution within the band; there are limited components outside this band. There are aliased peaks at higher frequencies, around twice the central frequency and above, but the amplitude of these higher frequencies is much lower than the main Fourier peak shown in Figure 2. 

\begin{figure}[htbp]\label{Fig_2}
	\centering
		\includegraphics[height=0.7\hsize]{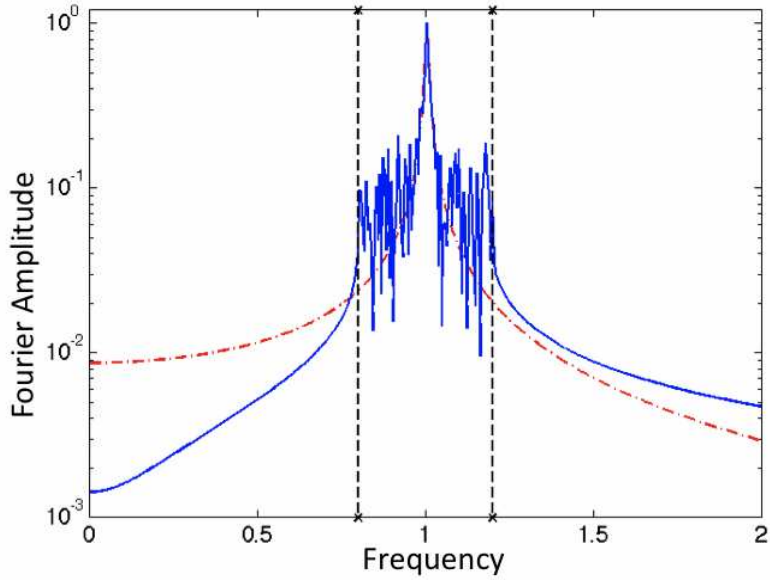}
	\caption{\label{fig:Fig2} (Color online) The frequency distribution for a reconstructed signal using the interleaved measurements (blue-solid line), with fitted resonance curve (red-dot-dashed line) and the sampling frequency band ($B=0.4$, black-dashed lines).}
\end{figure}
\begin{figure}[htbp]\label{Fig_3}
	\centering
		\includegraphics[width=0.9\hsize]{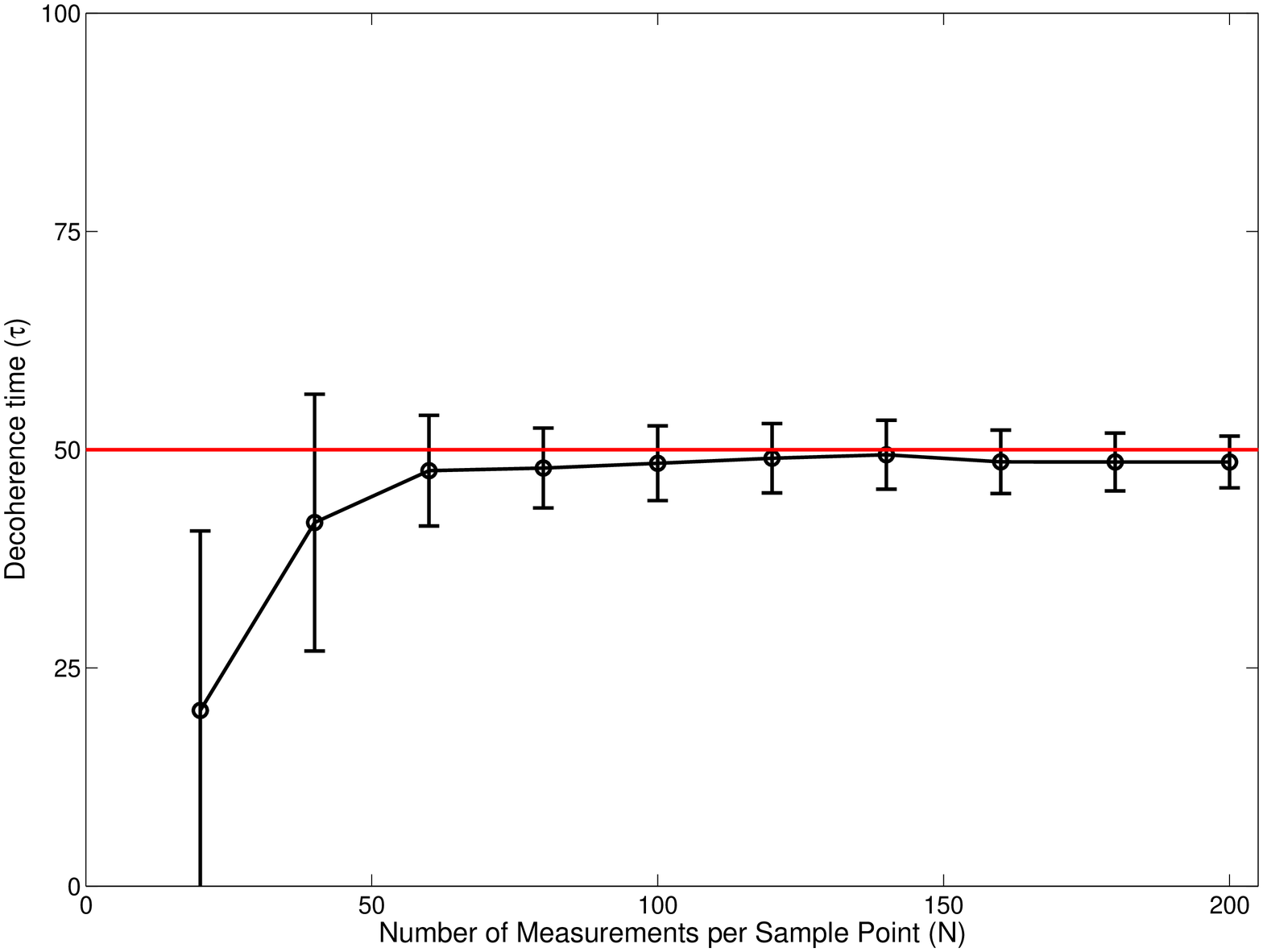}
	\caption{\label{fig:Fig3} (Color online) Estimates for the decoherence time from the fitted resonance ($\tau = 1/\kappa$) for different number of projective measurements per sample point: $f=1$, $\kappa = 0.02$, $B=0.4$ and $200$ oscillation cycles, mean values and errors. The error bars are $\pm\sigma$.}
\end{figure}
\begin{figure}[htbp]\label{Fig_4}
	\centering
		\includegraphics[width=0.9\hsize]{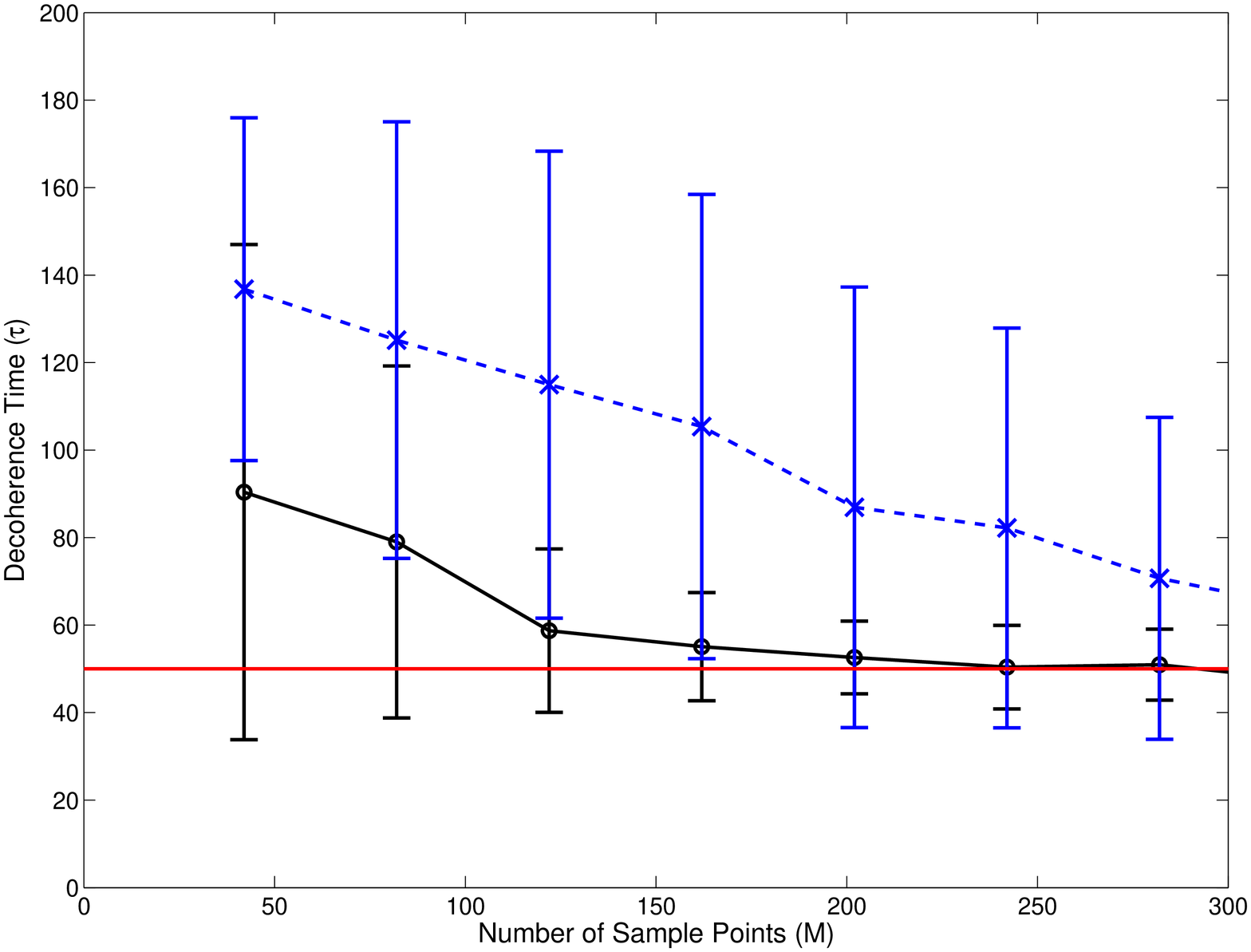}
	\caption{\label{fig:Fig4} (Color online) Convergence of the estimated decoherence time ($\tau = 1/\kappa$) as a function of the number of samples $M$ for the sinc method (blue-dash) and the interleaved method (black-solid). All points use $N=100$ and $f=1$, $\kappa= 0.05$, $B=0.4$ . The error bars are $\pm\sigma$.}
\end{figure}

Both the central frequency and the width of the resonance (e.g. the full peak half maximum) can be found by fitting a resonance curve to the Fourier transform of the reconstructed signal. This provides an estimate of the central frequency (and hence the Hamiltonian) and the decoherence rate, which is related to the width of the peak. The accuracy of the frequency estimate is dependent on the spacing of the Fourier coefficients and can be improved slightly by interpolating between Fourier coefficients~\cite{Huang00}. For the cases considered here the frequency errors are less than about 1\% and the main errors come in the estimation of the decoherence rate. The resonance curve corresponding to the coherent oscillations given in (\ref{me3}) is fitted to the signal peak in Figure 2 using a simple enumerative search through a range of parameter values (for frequency, $\kappa$ or $\tau$ and the peak amplitude). 

The average decoherence time (as given by $\tau = 1/\kappa$) is plotted (with error bars) in Figure 3 against the number of measurements ($N$) for each sample point. The mean value contains a statistical bias which underestimates the decoherence time (overestimating the environmental coupling) because the effect of aliasing is to increase the size of the Fourier coefficients in the tails of the resonance peak, which broadens the measured peak. If more measurements are taken for each sample point, the representation of the coherent oscillations will be better, the level of the noise in the frequency domain will be reduced and the frequency components outside the band of interest will be smaller, thereby reducing the effect of aliasing.  The minimum time required to take the measurements to characterize the qubit is only linear in the number of measurements at each sample point $N$, and Figure 3 shows that $N > 50$ gives a reasonable accuracy for the decoherence time. 

Figure~4 demonstrates the superiority of the interleaving method over the sinc construction. For $M > 150$ the interleaved estimate of $\tau$ provides a good estimate ($\tau = \kappa^{-1} \approx 50$) while the sinc method requires  significantly more samples ($M \gg 300$) to achieve similar results. Similar results were obtained for a large range of values for $N,M$ and $\kappa$ with $f=1$ fixed.

\section{\label{sec:sec5}CONCLUSIONS}
In this paper we have demonstrated that a sample (and time) efficient strategy exists for using a series of projective measurements to estimate the frequency of the coherent oscillations of a qubit subject to a decohering environment. This strategy uses an interleaved sampling technique from classical signal processing for the reconstruction of band-limited signals.  If the qubit resonance is confined to a narrow frequency band, this technique could offer significant savings in the numbers of measurements that are required to characterize the system and in terms of the length of time required to take the measurements. We have used a simulated example to generate sets of sample points that demonstrate that the technique is robust even when the coherent oscillations are not strictly confined to the relevant frequency band and when there are errors in the sample points themselves.

\begin{acknowledgements}
The authors would like to thank Andrew Greentree, Charles Hill and Lloyd Hollenberg at the University of Melbourne for helpful discussions during the preparation of this paper. JFR would also like to thank the Centre for Quantum Dynamics at Griffith University and the Department of Physics at the University of Melbourne for their hospitality.
\end{acknowledgements}

\bibliographystyle{unsrt} 
\bibliography{report2}

\begin{thebibliography}{10}

\bibitem{Nakamura99}
Y.~Nakamura, Yu.~A. Pashkin, and J.~S. Tsai.
\newblock Coherent control of macroscopic quantum states in a
  single-cooper-pair box.
\newblock {\em Nature}, 398:786--788, 1999.

\bibitem{Pashkin03a}
Yu.~A. Pashkin, T.~Yamamoto, O.~Astafiev, Y.~Nakamura, D.~V. Averin, and J.~S.
  Tsai.
\newblock Quantum oscillations in two coupled charge qubits.
\newblock {\em Nature}, 421:823--826, 2003.

\bibitem{Duty04}
T.~Duty, D.~Gunnarsson, K.~Bladh, and P.~Delsing.
\newblock Coherent dynamics of a josephson charge qubit.
\newblock {\em Phys. Rev. B}, 69:140503, 2004.

\bibitem{Schirmer04}
Sonia~G. Schirmer, A.~Kolli, and Daniel K.~L. Oi.
\newblock {\em Phys. Rev. A}, 69:050306(R), 2004.

\bibitem{Cole05}
Jared~H. Cole, Sonia~G. Schirmer, Andrew~D. Greentree, Cameron~J. Wellard,
  Daniel K.~L. Oi, and Lloyd C.~L. Hollenberg.
\newblock Identifying an experimental two-state hamiltonian to arbitrary
  accuracy.
\newblock {\em Phys. Rev. A}, 71:062312, 2005.

\bibitem{Cole06}
Jared~H. Cole, Andrew~D. Greentree, Daniel K.~L. Oi, Sonia~G. Schirmer,
  Cameron~J. Wellard, and Lloyd C.~L. Hollenberg.
\newblock Identifying a two-state hamiltonian in the presence of decoherence.
\newblock {\em Phys. Rev. A}, 73:062333, 2006.

\bibitem{Jerri77}
Abdul~J. Jerri.
\newblock The shannon sampling theorem - its various extensions and
  applications: a tutorial review.
\newblock {\em Proc. of the IEEE}, 65:1565--1596, 1977.

\bibitem{Lathi98}
B.~P. Lathi.
\newblock {\em Modern Digital and Analog Communication Systems, 3rd Ed.}
\newblock Oxford, 1998.

\bibitem{Kohlenberg}
A.~Kohlenberg.
\newblock Exact interpolation of band-limited functions.
\newblock {\em J. Appl. Phys.}, 24:1432--1436, 1953.

\bibitem{Vaughan91}
Rodney~G. Vaughan, Neil~L. Scott, and D.~Rod White.
\newblock The theory of bandpass sampling.
\newblock {\em IEEE Trans.\ Signal Proc.}, 39:1973--1984, 1991.

\bibitem{Ralph09}
J.~F. Ralph, K.~Jacobs, and C.~D. Hill.
\newblock Frequency tracking and parameter estimation for robust quantum
  state-estimation.
\newblock {\em Eprint: quant-ph/0907.5034}, 2009.

\bibitem{Gambetta01}
Jay Gambetta and H.~M. Wiseman.
\newblock State and dynamical parameter estimation for open quantum systems.
\newblock {\em {Phys.\ Rev.\ A}}, 64:042105, 2001.

\bibitem{WisMil10}
H.~M. Wiseman and G.~J. Milburn.
\newblock {\em Quantum Measurement and Control}.
\newblock Cambridge University Press, 2010.

\bibitem{Huang00}
D.~Huang.
\newblock Approximate maximum likelihood method for frequency estimation.
\newblock {\em Stat. Sinica}, 10:157--171, 2000.

\end{thebibliography}

\end{document}